\documentclass[12pt]{iopart}

\usepackage{graphicx}

\def\eion{{(e~+~ion)}\ }

\def\fexvii{{\rm Fe~\sc xvii}\ }
\def\fexviii{{\rm Fe~\sc xviii}\ }
\def\fexix{{\rm Fe~\sc xix}\ }
\def\fexx{{\rm Fe~\sc xx}\ }

\def\en{{$n$\ }}
\def\el{{$l$\ }}
\def\ii{{$i$\ }}
\def\dne{{$N_e$\ }}

\newcommand{\be}{\begin{equation}}
\newcommand{\ee}{\end{equation}}
\begin{document}

\title[RMOP-III.Plasma broadeneing]{R-matrix calculations for
opacities:~III. Plasma broadening of autoionizing resonances}

\author{A K Pradhan$^{1,2,3}$}

\address{$^1$ Department of Astronomy, $^2$ Chemical Physics Program, 
$^3$ Biophysics Graduate Program,
Ohio State University, Columbus, Ohio 43210, USA}
\vspace{10pt}

\begin{abstract}
A general formulation is employed to study and quantitatively 
ascertain the effect of plasma broadening of {\it intrinsic}
autoionizing (AI) resonances in photoionization cross sections.
In particular, R-matrix data for
iron ions described in the previous paper in the RMOP
series (RMOP-II, hereafter RMOP2) are used to demonstrate underlying physical
mechanisms due to electron collisions, ion microfields (Stark),
thermal Doppler effects, core
excitations, and free-free transitions. Breit-Pauli R-matrix (BPRM)
cross section for the large number of bound levels of Fe ions are
considered, 454 levels of Fe~XVII, 1,184 levels of Fe~XVIII and 508
levels of Fe~XIX. Following a description of theoretical and
computational methods, a sample of results is presented to 
show significant broadening and shifting
of AI resonances due to {\it Extrinsic} plasma broadening as a function
of temperature and density.
Redistribution of AI resonance strengths broadly preserves their
integrated strengths as well as the
naturally {\it intrinsic} asymmetric shapes of resonance complexes which
are broadened, smeared and flattened, eventually dissolving into the
bound-free continua. 
\end{abstract}

%
%
%
%
%

\section{Introduction}
 Resonances arise in most atomic interactions. They are especially
important in processes 
such as \eion scattering and photoionization. At the same time, 
plasma perturbations markedly affect atomic spectra susceptible to
varying temperature, density, and other factors.
 Whereas a vast body of literature exists on line broadening in
laboratory and astrophysics plasma
environments
\cite{g05,p81,k99,dk09,op,d06}, 
there is relatively little work on systematic
theoretical treatment of autoionizing resonances that are more
readily susceptible to plasma interactions \cite{k85,x12}, though
results have
been obtained for K-shell spectra (viz. \cite{pp16}) observed 
astrophysically \cite{h20}. 
Stark broadening and other broadening mechanisms for plasmas have been reviewed
from the perspective of individual lines and spectrum \cite{gig14,cs82}, 
and in non-local-thermodynamic-equilibrium \cite{hm15}.
However, opacity calculations require a statistical treatment such as
implemented in the Opacity Project (hereafter OP \cite{adoc13,nb2,op,mhd}).

Resonances are
ubiquitous in cross sections, measured and calculated in
a variety of ways with ever-increasing precision and resolution.
State-of-the-art experimental devices such as synchrotron based 
ion storage rings and narrowband 
photon sources can now resolve resonances in many atomic
systems. Coupled-channel calculations, mainly
using the R-Matrix method, have been carried out for nearly all elements
and ions up to at least iron under OP \cite{op,symp} and, more
extensively, the Iron Project (hereafter IP \cite{ip}). 
A prime feature of these calculations is the
presence of resonances all throughout the energy ranges of interest.
However, resonances are of different types, and exhibit varying shapes,
sizes and heights. Their overall resonance strengths may also be computed
in analogy with line oscillator strengths for modeling of radiative
processes \cite{aas}.

 But the question remains: how are resonance profiles affected by plasma
perturbations? To be more precise, how would the {\it intrinsic}
autoionization shape be modified by {\it extrinsic} particle
interactions in a given environment? 
The complexity of the problem
becomes evident when one considers that autoionization profiles are
inherently asymmetric, described by the Fano formula for isolated
resonances in terms of an asymmetry parameter and energy \cite{fano}.
But, any singular expression is insufficient to treat infinite overlapping
series of AI resonances which, in fact,
range from extremely narrow
Rydberg resonances approaching series limits, to huge
photoexcitation-of-core resonances that span hundreds of eV in energy
and considerably alter the background continuum
below core excitation threshold \cite{aas,n11}. 
Previous works and conventional approach to plasma modeling of
resonances, and collisional-radiative models, 
generally follow the
'isolated resonance approximation', which treats autoionizing resonances
as discrete bound levels and entail the calculation of the oscillator
strength at a single energy, followed by a perturbative plasma
broadening treatment based
on independently calculated autoionization and radiative rates
(viz. the Cowan code \cite{cowan}). Although a physical explanation is
lacking, arbitrarily increasing line
broadening factors of all lines by up to a factor up to $\sim$100
in atomic structure calculations is found to recover missing solar opacity
quantitatively \cite{k16}.

Ideally, what is needed is a theoretical method that can be translated
into a computational algorithm taking
into account the variety of resonance shapes and their positions
relative to the excited ion core level. 
Electron-ion interactions in a plasma lead to dominant
forms of broadening: Doppler, Stark and electron impact. The Doppler
width is approximated by Gaussian that is more narrowly peaked around
the line center, and falls off faster, than the other Lorentzian
profiles due to Stark and electron impact. The Stark effect due to ions
is particularly important for hydrogenic systems when it is linear due to
\el-degeneracy; a static approximation
is sometimes employed since ions move much 
slower than electrons \cite{g05,hm14}.
In contrast, the electron
impact broadening profile is a Lorentzian with much wider
effect on the line wings, and as the electron density and the temperature
of the plasma increases, electron collisions become the dominant source
of broadening. That would especially be the case for weakly bound
electrons in doubly-excited autoionizing states, which would be perturbed
more than bound electrons considered in line broadening theories.

In this paper we present a computational methodology that
aims to incorporate electron impact broadening in a generally applicable
manner suitable for laboratory and astrophysical plasma sources.
Without loss of generality, and based on large-scale coupled channel
R-matrix calculations (\cite{b11,n11}), 
we consider the photoionization of a complex
atomic system, neon-like to fluorine-like iron, 
\fexvii $\longrightarrow$ \fexviii, 
in this study as exemplar of applicability to atomic processes in
plasmas. 

\section{Theoretical formulation} 

 We first sketch out the theoretical outline for channel coupling that
gives rise to resonances and then the resonance broadening modeled after
line broadening due to electron impact.

\subsection{Resonances and channel coupling}

 Autoionizing resonances manifest themselves via inter-channel coupling
in the coupled channel (CC) framework. 
In the CC approximation
the atomic system is represented as the 'target' or the 'core' ion of
N-electrons interacting with the (N+1)$^{th}$ electron. The (N+1)$^{th}$
electron may be bound in the electron-ion system, or in the electron-ion
continuum depending on its energy to be negative or positive. The total
wavefunction, $\Psi_E$, of the (N+1)-electron system in a symmetry
$J\pi$ is an expansion over the eigenfunctions of the target
ion, $\chi_{i}$ in specific state $S_iL_i(J_i)\pi_i$, coupled with the
(N+1)$^{th}$ electron function, $\theta_{i}$:
\begin{equation}
\Psi_E(e+ion) = A \sum_{i} \chi_{i}(ion)\theta_{i} + \sum_{j} c_{j}
\Phi_{j},
\end{equation}
where the $\sum_i$ is over the ground and excited states of the target or the
core ion. The (N+1)$^{th}$ electron with energy $k_{i}^{2}$
corresponds to a channel labeled
$S_iL_i(J_i)\pi_ik_{i}^{2}\ell_i(SL(J)\pi)$.
The $\Phi_j$s are bound channel functions of the (N+1)-electron system
that account for short range correlation not considered in the first
term and the orthogonality between the continuum and the bound electron
orbitals of the target.

 Depending upon the total energy $E$ of the \eion system, and the channel
energy $k_{i}^{2} > 0$ or $k_{i}^{2} < 0$, a channel may be
open or closed relative to an ion level $E_i$. 
Inter-channel interactions between open and closed channel
wavefunctions result in resonances below the excitation threshold at
$E_i$. If $E < 0$ for all channels then the \eion system is in a pure
bound state; otherwise we have a free state with an electron in the
continuum and some channels open and some closed. Therefore, the CC
wavefunction expansion Eq.~(1) may be used to obtain either \eion
collision strengths or bound-bound and bound-free radiative parameters
such as oscillator strengths and photoionization cross sections.

 With reference to Fig.~1, we have the position of a given resonance
$\omega_r$
corresponding to an excitation threshold $E_i$ in terms of its effective
quantum number $\nu_i$ as 

\be \omega_r = \omega_g + E_i -\frac{(z+1)^2}{\nu_i^2}. \ee 

 That yields

\be \nu_i(\omega_r) = \left[ \frac{(z+1)^2}{\omega_g + E_i -
\omega_r]}\right]^{1/2}. \ee  

Typically, there are many excited levels $E_i$ included
in coupled channel calculations and may number in the hundreds.
Infinite series of resonances $E_i \nu_{n \ell}$ arise and
converge on to each level $E_i$. There can be
considerable overlap between 
 weakly bound narrow high-$\nu$ Rydberg resonances converging on to
and immediately below a given threshold, and deeply bound strong and wide
resonances with low $\nu$-values belonging to higher levels. 
A computational algorithm must successively 
convolve groups of resonances
identified with respect to all ion core levels.

Let $\tilde{\sigma}(\omega')$ be 
 the computed cross section and $\sigma(\omega)$ the convolved cross
section such that

\be \sigma(\omega) = \int \tilde{\sigma}(\omega') \phi(\omega,\omega') 
d \omega' , \ee

where the profile factor is

\be \phi(\omega,\omega') = \frac{\gamma(\omega)/\pi}{(\omega-\omega')^2 +
\gamma(\omega)^2}. \ee 

\subsection{Resonance broadening mechanisms}

A general theoretical approximation
for scattering of a free electron with an electron in
doubly-excited quasi-bound states 
is necessarily computationally intensive since it needs to be
incorporated within a coupled channel framework, and superimposed on
{\it ab initio} calculations of cross sections.
Primary broadening mechanisms such as electron collisions,
Stark broadening due to
ion microfields, and
Doppler broadening due to thermal motions need to considered {\it a
priori}. We develop a theoretical treatment that
accounts for these physical effects independently within a computational
viable procedure. 

 The parameters in the formulation are derived in analogy with 
line broadening but modified significantly to apply to AI resonances. 
In the present formulation we associate the energy to the effective
quantum number relative to each threshold $\omega' \rightarrow \nu_i$ to
write the total width as:

\begin{eqnarray}
\gamma_i(\omega,\nu,T,N_e) &  = & \gamma_c(i,\nu,\nu_c)+
\gamma_s(\nu_i,\nu_s^*)\\
 & + &  \gamma_d(A,\omega) + \gamma_f(f-f;\nu_i,\nu_i'), \nonumber 
\end{eqnarray}

pertaining to
collisional $\gamma_c$, Stark $\gamma_s$, Doppler $\gamma_d$, and
free-free transition $\gamma_f$ widths
respectively, with additional parameters as defined below.
Without loss of generality we assume a Lorentizan profile
factor that describes collisional-ion broadening which dominates in
HED plasmas. We assume this approximation to be valid since
collisional profile wings extend much wider as $x^{-2}$, compared to
the shorter range $exp(-x^2)$ for thermal Doppler, and $x^{-5/2}$ for
Stark
broadening (viz. \cite{adoc13}). In principle the limits of integration
in Eqs. (4-6) are $\mp \infty$, which are
replaced in practical calculations 
by $\mp \gamma_i/\sqrt{\delta}$, where $\delta$ is
chosen to ensure full Lorentzian profile energy range
and for accurate normalization.
Convolution by evaluation of Eqs. (3-6) is carried out for each
energy $\omega$ throughout the tabulated mesh of energies used to
delineate all AI resonance structures, for each cross section,
and each core ion threshold. 

\subsubsection{Electron impact broadening}
 At sufficiently high densities collisional broadening is dominant and
and mathematically represented by a Lorentizan function (Eq.~5) that
correctly approximates the slowly varying behaviour in the line wings. 
 We develop a numerical procedure for convolving cross sections
including resonances over a Lorentzian damping width.
Given energy dependent cross sections tabulated at sufficiently fine
mesh, we first switch the energy variable to the effective quantum number
$\nu = z/\sqrt(E)$, where $E = \hbar \omega$. In photoionization, we
take $\omega$ to be the photon frequency; henceforth we shall 
also employ $\omega$ as the energy variable assuming atomic units 
$\hbar$ = 1. 
The $\nu$ is more appropriate since for a
resonance it is defined relative to the excited core ion level, as
illustrated in Fig.~1. 

\begin{figure*}
{ \includegraphics[width=3.0in,height=3.0in]{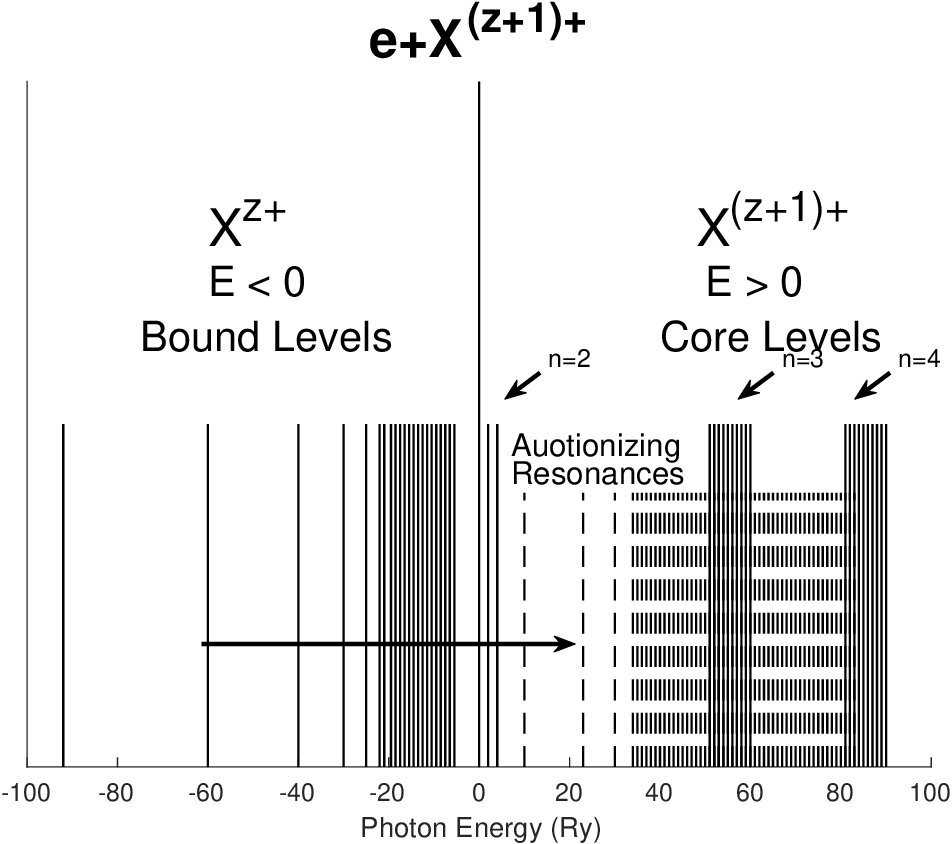}}
{ \includegraphics[width=3.0in,height=3.0in]{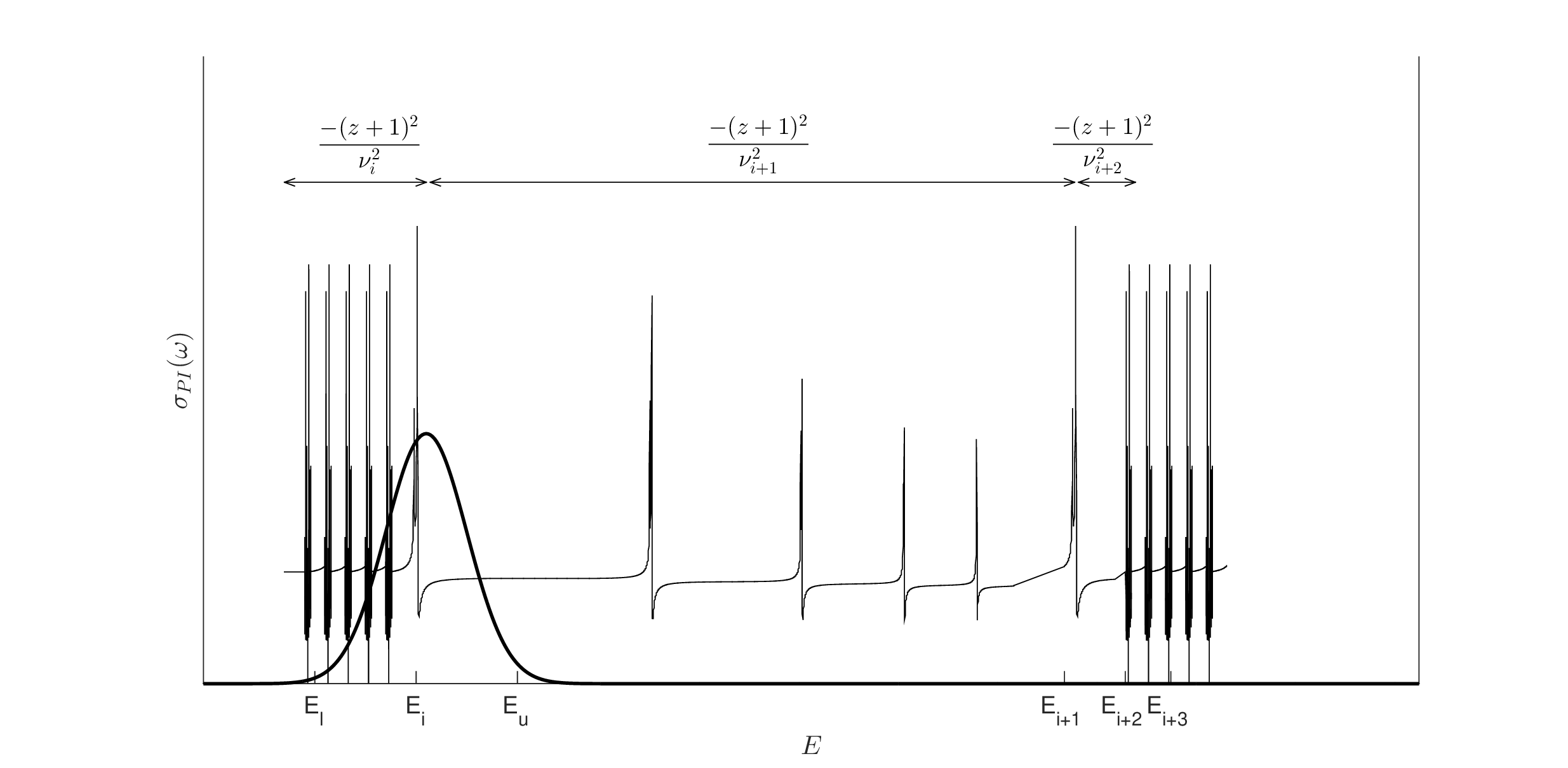}}
\caption{{\bf Left:} Schematic diagram of a coupled channel calculation for
photoionization of bound states (solid lines) of an 
ion $X^{z+} \rightarrow X^{z+1}$ --- AI resonances (dashed lines) correspond
to Rydberg series converging on to excited levels of the residual ion
with $E = -(z+1)^2/\nu^2$; {\bf Right:} ion thresholds of convergence 
$E_i, \ E_{i+1},
\ E_{i+2}, \ E_{i+3}....$ and a 
 Lorentzian profile with lower and upper energy limits ($E_\ell, E_u$)
spanning narrow high-$n$ resonances below $E_i$ and broader ones above.} 
\end{figure*}

We consider photoionization of an ion of element X 
with charge $z$ in an initial state by photon of energy $\hbar \omega$
 into the ground or excited level of a residual ion of 
charge $(z+1)$

\be \hbar \omega + X^{z+} \longrightarrow e + X^{z+1}. \ee

 It is assumed that unperturbed photoionization cross sections
$\tilde{\sigma}_{PI}(\hbar \omega)$ are
theoretically computed with sufficient resolution in energy to delineate
autoionization profiles. According to the impact approximation
\cite{op} we may
then represent the damping profile with a Lorentzian expression

\be \phi_\omega(E) = \frac{\gamma/\pi}{[(E + x - E_o)^2 +
(\gamma^2]}. \ee

 In analogy with electron impact damping of bound-bound line
transitions, we define
$E_o$ as the resonance center, $\gamma$ as the width and $x$ the 
energy shift (later we shall assume that $|E - E_o| >> x$). 
We may further express

\be N_e \gamma = \gamma + ix, \ee

 where N$_e$ is the electron density and $\gamma$ is the damping
constant which may be written in terms of the electron
distribution $f(\epsilon,T)$ at a given temperature T as

\be \gamma(T) = \int_0^{\infty} v Q_D(\epsilon) f(\epsilon,T) d\epsilon.
\ee 
 
 Given Q$_D$ as the electron impact cross section and a Maxwellian
distribution we may obtain the thermally averaged damping rate coefficient 

\be \Upsilon_D(T) = \int_0^{\infty} \Omega_D(\epsilon) exp(-\epsilon/kT)
d(\epsilon/kT), \ee

where $\Omega(\epsilon)$ is the collision strength.  Then

\be \gamma(T) = 2\frac{\hbar^2}{m} \left(\frac{\pi}{mkT}\right)^{1/2} \Upsilon_D
\ee.

 In Eqs. (8-12) the $\Upsilon_D$ is a complex quantity. However, 
for small $\delta \omega = (\omega - \omega_o)$ in the one-perturber
approximation (\cite{op} and references therein), 
we have $\gamma = N_e \gamma$ and
$\phi_\omega = (\gamma/2\pi)/(\omega-\omega_o)^2$.

Now we establish a correspondence between $\gamma(\omega)$ and the
electron impact rate coefficient $\Upsilon$ according to the relation

\be \gamma(\omega) = 2 \left(\frac{\pi}{kT}\right)^{1/2} a_o^3 N_e \Upsilon(\nu), \ee

where $\Upsilon(\nu)$ is computed at the resonance energy 
corresponding to $\nu = z/\sqrt(E)$, with E in Rydbergs and atomic
units $a_o = \hbar = 1$. We now approximate

\be \Upsilon(\nu) \approx G(z) <r_\nu^2> = G(z) \frac{5\nu^4}{2(z+1)^2}. 
\ee

 G(z) is an effective Gaunt factor for electron impact excitation of
positive ions, empirically determined for line broadening work in
OP \cite{op} to be 

\be G(z) = 6.3 - \frac{5.9}{(z+1)}. \ee

 The behavior of G(z) with ion charge $z$ and temperature T 
has been further studied for electron impact broadening, and we adopt an
improved expression (\cite{adoc13,dk81,x12})

\be G(T,z,\nu_i) = \sqrt 3/\pi [1/2+ln(\nu_i kT/z)]. \ee

For example, in Table~\ref{tab:t1} we compare the two expressions 
and find that they differ
significantly for $\nu < 10$, but $G(T,z,\nu \rightarrow G(z)$ as $\nu
\rightarrow 10$, and exceeds marginally for $\nu > 10$ when BPRM
resonance structure calculations are truncated. 

\begin{table}
\caption{Gaunt factor for electron impact collisional broadening:
dependence on temperature T(K), ion charge $z$ and effective quantum
number $\nu$ of
excited levels (Eq.~16) $T = 2\times10^6K, \ z = 16$.
\label{tab:t1}}
\begin{center}
\begin{tabular}{|c|c|}
\hline
$\nu$ & G(T,z,$\nu$) \\
\hline
 3.0 & 1.75\\
 4.0 & 2.52\\
 5.0 & 3.12\\
 6.0 & 3.60\\
 7.0 & 4.02\\
 8.0 & 4.37\\
 9.0 & 4.69\\
 10.0 & 4.97\\
\hline
\end{tabular}
\end{center}
\end{table}

 Here $\omega_g$ is the ionization energy of the ground state of the
photoionizing ion X$^{z+}$. Then from Eq.~(18) we obtain the
temperature-density dependent width at each energy

\be \gamma_i(\omega_r; Ne, T) = 5 \left( \frac{\pi}{kT} \right) ^{1/2} a_o^3 N_e
G(z) \frac{\nu_i^4(\omega_r)}{(z+1)^2} . \ee

 Evaluating the constants with T(K) and $N_e$ cm($^{-3}$), we obtain 

\be \gamma_i(\omega_r; Ne, T) = 5.2184 \times 10^{-22} \left( \frac{N_e}{T^{1/2}}\right)
\left( \frac{G(z)}{(z+1)^2}\right) \nu_i^4(\omega_r).\ee

With the
transformation of the unbroadened cross section using Eq.~(18), 

\be \tilde{\sigma}(\omega) \longrightarrow \sigma(\omega; T, N_e), \ee

we obtain the temperature-density-energy dependent functional representing the
photoionization cross section broadened by electron impact. This greatly
expands the scope of the calculations since  Eq.~(19)
implies that the convolution must be carried out at each
energy in the tabulated energy mesh (transposed as $E(\omega) \rightarrow \nu)$
of unbroadened function $\tilde{\sigma}(\omega)$, with another
tabulation for the Lorentzian profile Eq.~(8), and for each temperature
and electron density. In the next section we describe the procedure
developed for such numerical calculations.

Given $N$ core ion levels corresponding to resonance
structures,

\be \sigma(\omega) = \sum_i^N \left[ \int \tilde{\sigma}(\omega')
\left[ \frac{\gamma_i(\omega)/\pi}{x^2 +
\gamma_i^(\omega)}\right] d \omega' \right]
. \ee

 With $x \equiv \omega' - \omega $, the summation is over all excited
thresholds $E_i$ included in
the $N$-level CC or RM wavefunction expansion, and corresponding
to total damping width $\gamma_i$ due to all broadening processes.
The profile $\phi(\omega',\omega)$ is centered at each
continuum energy $\omega$, convolved over the variable $\omega'$ and
relative to each excited core ion threshold \ii.

We employ the following expressions for
computations:

\be \gamma_c(i,\nu) \  = \ 5 \left( \frac{\pi}{kT} \right)^{1/2}
 a_o^3 N_e G(T,z,\nu_i) (\nu_i^4/z^2), \ee

where T, \dne, $z$, and $A$ are the temperature, electron density, ion
charge and atomic weight respectively, and $\nu_i$ is the effective
quantum
number relative to each core ion threshold \ii: $\omega \equiv E =
E_i-\nu_i^2/z^2$ is a continuous variable. 
A factor
$(n_x/n_g)^4$ is introduced for $\gamma_c$
to allow for doubly excited AI levels with excited core
levels $n_x$ relative to the ground configuration $n_g$
(e.g. for \fexviii
$n_x=3,4$ relative to the ground configuration $n_g=2$).

\subsubsection{Stark broadening}

A treatment of the Stark effect for complex
systems entails two approaches, one where both electron and ion
perturbations are combined (viz. \cite{dk87,x12}), or separately (viz.
\cite{op,adoc13}) employed herein. Excited Rydberg levels are nearly
hydrogenic
and ion perturbations are the main broadening effect, though collisional
broadening competes significantly increasing with density
as well as $\nu_i^4$ (Eq.~14).
For bound levels in a plasma microfield of strength F, the Stark sub-levels 
of a level $n$ span a range given by the highest component ($n,k_{max}$) 
with energy (viz. \cite{op,adoc13})

\be E(n,k_{max}) = -\frac{z^2}{n^2} + \frac{3}{z} n(n-1) F \ee
and the lowest component of sub-level ($(n+1),k_{min}$) with energy
\be E(n+1,k_{min}) = -\frac{z^2}{(n+1)^2} - \frac{3}{z}n(n+1)F. \ee  

In deriving occupation probabilities in the Mihalas-Hummer-D\"{a}ppen
equation-of-state (MHD-EOS) \cite{mhd} used in OP
work \cite{op}, a critical field strength $F_c$ is calculated 
when Stark broadening renders these
two components equal, and Stark ionization dissolves level $n$
into the continuum.
The total Stark width of a given \en-complex is $\approx (3F/z)n^2$.
Assuming the dominant ion perturbers to be protons and density
equal to electrons, \dne=$N_p$, and replacing $n$ by the effective
quantum number $\nu_i$ relative to each excited threshold of an ion with
charge $z$, we take $F=[(4/3)
\pi a_o^3 N_e)]^{2/3}$, as employed in MHD-EOS for 
Stark broadening in Eq.~(6)

\be \gamma_s(\nu_i,\nu_s^*) =
[(4/3)\pi a_o^3 N_e]^{2/3} \nu_i^2. \ee

In addition, in employing Eq. (6) a Stark ionization parameter
$\nu_s^* = 1.2\times 10^3 N_e^{-2/15}z^{3/5}$ is introduced such
that AI resonances may be considered fully dissolved into the continuum
for $\nu_i > \nu_s^*$, analogous 
to but distinct from the Inglis-Teller series limit \cite{it39}, or the
Stark ionization of {\it bound} (not AI) energy 
levels as considered in the MHD-EOS \cite{mhd}.

All calculations are carried out with and without
$\nu_s^*$ as shown later in Table~2, and illustrated in the
Figs.~\ref{fig:fe17}, \ref{fig:fe18}, \ref{fig:fe19}
presented herein (red and blue curves respectively). Results are
practically indistinguishable with and without Stark ionization cut-off,
and effect on redistribution of differential oscillator strength or
opacities. However, $\nu_s^*$ is a parameter that should prove to be
useful in further extension of plasma effects including Debye screening, as
discussed later.

\subsubsection{Thermal Doppler broadening}

The Doppler width is:

\be \gamma_d (A,T,\omega) = 4.2858 \times 10^{-7} \sqrt(T/A), \ee

where $\omega$ is {\em not} the usual line center but taken to be each
AI resonance energy. 

\subsubsection{Free-free transitions broadening}

The last term $\gamma_f$ in Eq. (6) accounts for
free-free
transitions among autoionizing levels with $\nu_i,\nu_i'$ such that 

\be X_i + e(E_i,\nu_i) \longrightarrow X_i' + e'(E_i',\nu_i'). \ee

The large number of free-free transition probabilities for $+ve$ energy
AI
levels $E_i,E_i' > 0$ may be computed using RM or atomic structure
codes (viz. \cite{s00,ss}). 
Free-free transitions are not
considered in the results in Figs.~2 and 3 but included in the
results discussed in Table 1, although it is found to be practically
negligible.

\section{Computational algorithm}

 In order to elicit and illustrate important physical features of the
formulation, we sketch a few salient features of the mathematical
algorithm developed to implement the procedure (numerical details and
the computer program will be presented elsewhere).

 We have re-defined the Lorentzian profile Eq.~(5) as in Eq.~(8), using a
damping rate coefficient Eqs.~(10-13) and Maxwellian electron distribution,
dependent on electron density and temperature as in Eqs.~(17-18).
Numerical evaluation scheme based on this formulation 
requires several practical considerations to be incorporated in the
computational algorithm and computer program.

\subsection{Profile limits}

 The limits of integration in Eq.~(4) are determined by the extent of the
Lorentzian factor in Eq.~(8). It needs to be ensured that the profile
extends into the resonance wings and/or approaches the background
continuum without loss of accuracy. Measuring the energy spread relative
to the resonance center $\omega = \omega_r$, we note that according to
Eq.~(13) $\omega = \omega_g + E_i$, with respect to the ionization potential and the target 
excitation energy $E_i$ above the ground state of the residual ion.
Then the profile maximum is (Eq.~8)

\be \phi_{max} (\omega = \omega_r) = \frac{1}{\pi \gamma(\omega)}. \ee

 We introduce an accuracy parameter $\delta$ and choose the profile
limits $\pm \omega_o$ such that

\be \phi (\omega = \omega_o) = \delta \phi_{max} = \frac{\delta}{\pi
\gamma(\omega)}.
\ee 

 Then,

\be \frac{\delta}{\pi \gamma(\omega)} =
\frac{\gamma(\omega)/\pi}{(\omega-\omega_o)^2 +
\gamma^2(\omega)}. \ee

 Or,

\be (\omega-\omega_o)^2 = \gamma^2(\omega) \left( \frac{1}{\delta} - 1 \right) . \ee

 For small $\delta$,

\be (\omega - \omega_o)^2 \approx \frac{\gamma^2(\omega)}{\delta}. \ee

 Therefore, $|\omega - \omega|$ limits the convolution
profile such that 

\be  \omega - \omega_o = \pm \frac{\gamma}{\delta^{1/2}}. \ee

 Whereas Eq.(4) using Eq.~(5) has an analytical solution in terms of
$tan^{-1}(x/\gamma)/\gamma$ evaluated at limiting values of $x
\rightarrow \mp \gamma/\sqrt\delta$, its evaluation
for practical applications entails piece-wise integration across
multiple energy ranges spanning many excited thresholds and different
boundary
conditions. For example, the total width $\gamma$ is very large at
high densities and the Lorentzian profile may be incomplete
above the ionization threshold and therefore not properly normalized.
We obtain the necessary redward left-wing correction for partial
renormalization as

\be
\lim_{a \rightarrow - \gamma/2\sqrt\delta}
\int_a^{+\gamma/\sqrt\delta} \phi(\omega,\omega') d\omega' =
\left[ \frac{1}{4} - \frac{tan^{-1}(\frac{a}{\gamma/2\sqrt\delta})}{\pi}
\right],
\ee

where $a$ is the lower energy range up to the ionization threshold,
reaching the maximum value $-\gamma/2\sqrt\delta$. The parameter
$\delta$ is generally chosen to be $10^{-2}$ so that the total profile
ranges over 10$\gamma$.

\subsection{Convolution quadrature} 

 The complexity of the problem arises from the following main factors:
(i) wide variety of narrow and broad
resonances, (ii) overlapping infinite Rydberg series
 belonging to a large number of excitation thresholds of the target ion,
and
(iii) Lorentzian profiles that vary at each energy on a mesh that is
independent of the tabulated energy mesh for the original
cross section. The schematics are described in Fig.~2. 

\begin{figure*}
\begin{center}
{ \includegraphics[width=3.5in,height=3.0in]{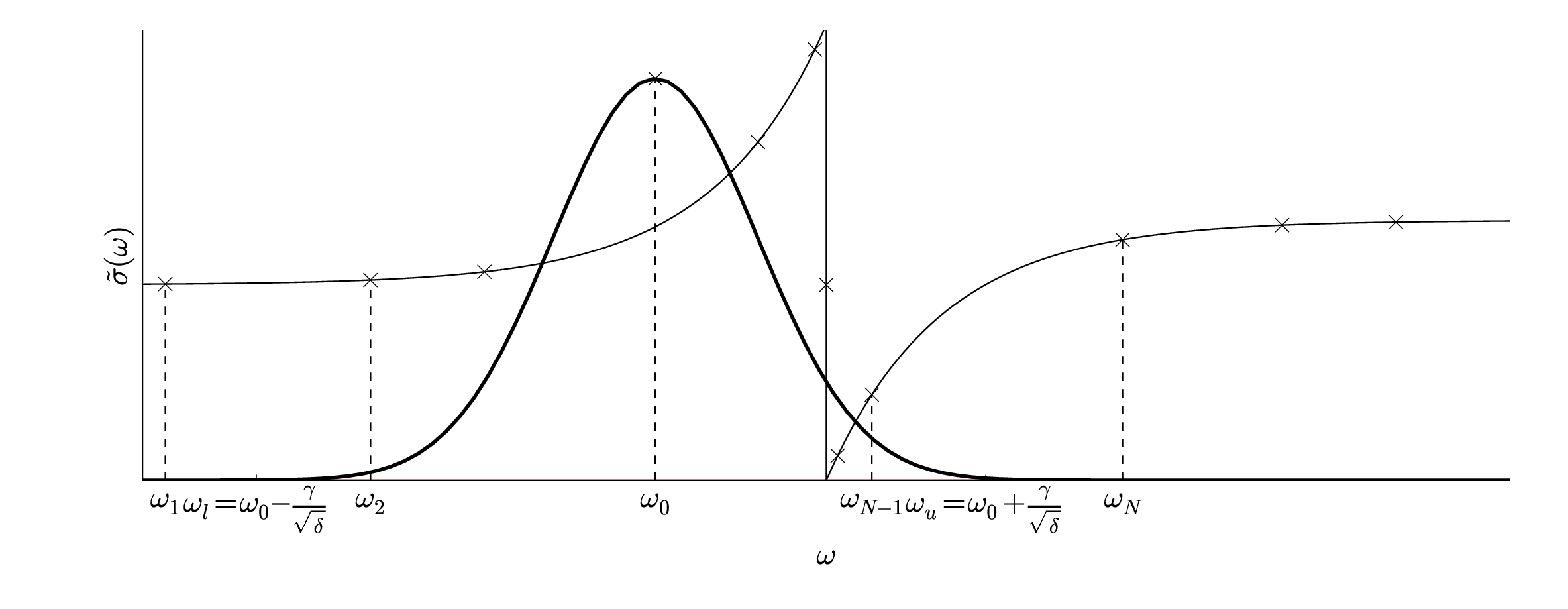}}
{ \includegraphics[width=2.5in,height=3.0in]{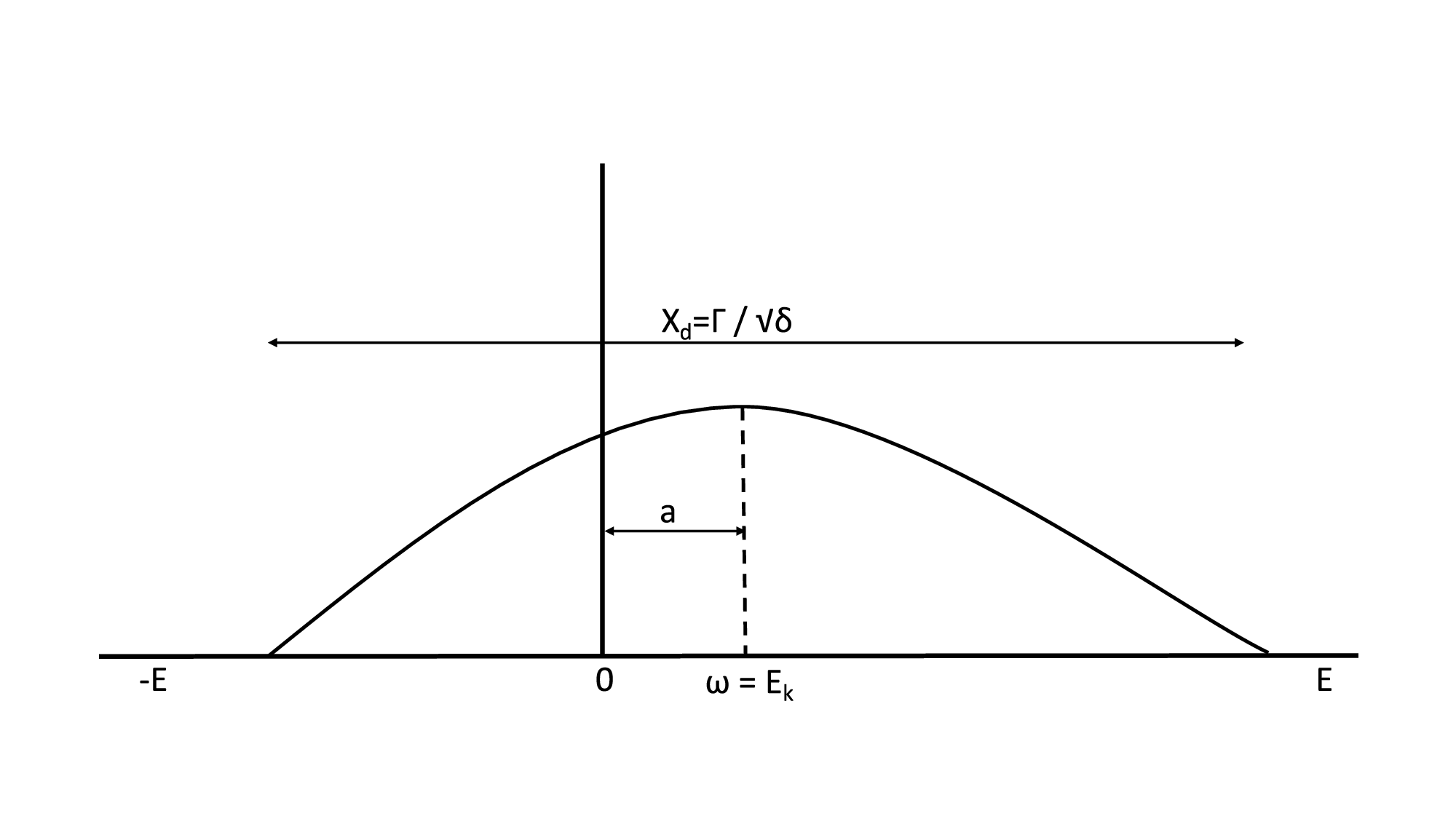}}
\end{center}
\caption{
{\bf Left:} Tabulated cross sections at energies ($hbar$ = 1) 
$\omega_1,.....\omega_N$ spanning a resonance centered at $\omega_o$, 
and Lorentzian profile with lower and upper limits $\omega_\ell = \omega_o -
\gamma/sqrt(\delta), \omega_u = \omega_o + \gamma/sqrt(\delta)$.
Point-by-point normalized profile convolution ensures a complete quadrature. 
{\bf Right:} Incomplete profile centered at $\omega = E_k$ 
with lower energy redward cut-off at ionization threshold on the left
and partial renormalization as in Eq.~31.}
\end{figure*}

 Numerically, we need to evaluate the integrand in Eq.~(20) using Eq.~(9), i.e.

\be \sigma(\omega) = \sum_i \left[ \int \tilde{\sigma}(\omega')
\left[ \frac{\gamma_i(\omega)/\pi}{(\omega-\omega')^2 +
\gamma_i^(\omega)}\right] d \omega' \right]
. \ee

 Here the summation is over all excitation thresholds $E_i$ included in
the CC wavefunction expansion (Eq.~1) and corresponding
damping widths $\gamma_i$. 
The profile $\phi(\omega',\omega)$ is centered at $\omega$; we
define $x \equiv \omega' - \omega $ (note change of order of variables
which is immaterial), then

\be \sigma(\omega) = \sum_i \left[ \frac{\gamma_i}{\pi}
\int_{-\frac{\gamma_i}{\sqrt{\delta}}}^{+\frac{\gamma_i}{\sqrt{\delta}}}
 \frac{\tilde{\sigma}(x)}{x^2 + \gamma_i^2} dx. \right] \ee
 
 This equation requires discrete summation over all target ion thresholds, and 
piecewise integration over normalized
profile at each energy. First, we consider the endpoints with lower
energy limit $x_\ell \equiv -(\omega_o - \omega) = -\gamma_i/\sqrt{\delta}$, and
upper limit  $x_u \equiv +(\omega_o - \omega) = +\gamma_i/\sqrt{\delta}$.
Let the tabulated energy mesh be $\omega_1, \omega_2, .....\omega_N$. 
Then $x_1 = \omega_1 -
\omega$, $x_2 = \omega_2 - \omega$, ........,$x_N = \omega_N - \omega$.
Assuming the lower limit $x_\ell$ to lie between 
${x_1 < x_\ell < x_2}$; and the upper limit $x_u$ between $x_{N-1} <
x_u < x_N$, we have

\begin{eqnarray}
\sigma(\omega) & = \sum_i \left[ \frac{\gamma_i}{\pi} \int_{x_\ell}^{x_2}
\frac{\tilde{\sigma}(x)}{x^2 + \gamma_i^2} dx + \int_{x_3}^{x_4} (...)
 dx +...\right] \\
 & +....+ \left[ \int_{x_{N-1}}^{x_N} (.....) dx \right].
\end{eqnarray}

\subsection{Interpolation and evaluation}

 Each of the raw originally tabulated unbroadened cross sections
$\tilde{\sigma}(\omega')$ needs to be
interpolated on to the resonance profile mesh. A linear interpolation is 
sufficient
for precision since the CC calculations are usually carried out at a
fine mesh to resolve most autoionizing 
resonances up to $\nu_i \leq = \nu_{max} = 10$ below each target
threshold $E_i$. Suppose the transposed energy mesh $\omega$ on to the resonance
profile is represented by linearly interpolated segments $a_j + b_jx$ with
$a_j,b_j$ coefficients such that,
$ x_\ell = -\gamma/\sqrt(\delta) < x < x_2 \longrightarrow
\sigma_1(\omega) = a_1 + b_1x$, $b_1 =
[\tilde{\sigma}(\omega_2)-\tilde{\sigma}(\omega_1)]/(\omega_2-\omega_1)$ ; 
$ x_2 < x < x_3 \longrightarrow
\sigma_2(\omega) = a_2 + b_2x$, $b_2 =
[\tilde{\sigma}(\omega_3)-\tilde{\sigma}(\omega_2)]/(\omega_3-\omega_2)$;.............
$ x_(N) < x < x_u = +\gamma/\sqrt(\delta) \longrightarrow
\sigma_N(\omega) = a_N + b_Nx$, $b_N =
[\tilde{\sigma}(\omega_N)-\tilde{\sigma}(\omega_{N-1}]/(\omega_N-\omega_{N-1})$
. Then for all thresholds ${i}$, 

\be \sigma(\omega) = \sum_i \frac{\gamma_i}{\pi} \left [\sigma_1(\omega)
+ \sigma_2(\omega)+..........+ \sigma_N(\omega) \right ].\ee 

 It is understood that the interpolation and summation is carried out with 
respect to profiles corresponding to all target ion thresholds at $E_i$.
 Having determined coefficients ${a_j,b_j}$ we need to evaluate
expressions for each segment as

\be \sigma_j(\omega) = \frac{\gamma_i}{\pi} \int_{x_j}^i{x^{j+1}} \frac{(a_j
+ b_jx)}{x^2 + \gamma^2} dx. \ee

 Evaluating separately,

\be \sigma_j(\omega) = a_j \left [\frac{tan^{-1}(x/\gamma_i)}{\gamma_i}
\left |_{x_j}^{x_{j+1}} \right ] + \frac{b_j}{2} \left [ ln (x^2 +
\gamma_i^2) \right |_{x_j}^{x_{j+1}} \right ]. \ee

 For clarity we have avoided the use of double scripts $(i,j)$,
one with respect to thresholds $E_i$ and the other for
interpolation between respective resonance profile segments. But in
principle we may represent the final values of the cross sections
convolved over all resonances at the transposed energy mesh $\omega'
\rightarrow \omega$ as

\be \sigma(\omega) = \sum_{i,j} \sigma_j^i (\omega), \ee

subsuming all target ion levels (Fig.~1 and Eq.~1) and interpolation into the
computational algorithm. Finally, we compute broadened cross sections at
the same energy mesh as the unbroadened cross sections
$\tilde{\sigma}(\omega')$ so that there is one-to-one correspondence
$\omega' \rightarrow \omega$. However, we note that the {\it
intermediate} energy mesh of the Lorentzian profile is independent, and
interpolated in accordance with the damping width Eq.~(11-12) at each
energy.  
 
\subsection{Computer program}

 A general program for convolving AI resonances
has been written and will be reported elsewhere. 
Here we note a few of the main features. The primary loops in the
program are over electron temperature $T_e$, density $N_e$, and target
thresholds $E_i$. The input is the unbroadened CC cross sections tabulated at a
sufficiently fine mesh to resolve resonances so that convolution,
interpolation and integration do not result in loss of accuracy. 
The accuracy parameter $\delta$ is chosen to be in the range $10^{-2} -
10^{-6}$; more importantly, it is ensured that the convolved cross
sections have converged, physically implying that the resonance wings have 
merged into the continuum.
The CPU time required depends mainly on the density which
determines the total width $\gamma$; for example, in the reported
calculations for \fexvii at T=$2\times10^6$K it is few minutes for 
\dne = $10^{21}$ cc and $\sim$3 hours for \dne=$10^{24}$cc.

 The program is suitable as a module within a post-processing program
for CC cross sections with AI resonances for photoionization,
electron-ion collisions and recombination,
intended for practical application in specified
temperature-density range.

\section{Results and discussion}

 The complexity and magnitude of RMOP computations has been studied
using photoionization data for a large number of bound 
levels of the three Fe ions described in RMOP2. Since AI plasma broadening
must be carried at each temperature-density pair, the resulting cross
sections constitute a huge amount of data required for opacities
calculations in HED plasma sources. In this section we discuss a 
small sample of results for those Fe ions to illustrate some 
physical features.

\subsection{\fexvii: Temperature-Density dependence}

Owing to its closed shell ground configuration and many excited
$n$-complexes of configurations, Ne-like \fexvii 
later works.
is of considerable importance in astrophysical and
laboratory
plasmas, as described in a number of previous works (\cite{np16} and
references therein).
The \fexvii BPRM calculations are carried out
with 218 fine structure levels dominated by $n=2,3,4$ levels of the core
ion \fexviii. The computed \fexvii
bound levels ($E<0$)
are dominated by configurations $1s^22s^22p^6 (^1S_0),
1s^22s^p2p^qn\ell, [SLJ] \ (p,q
= 0-2, \ n \leq 10, \ \ell \leq 9, \ J \leq 12$). The core \fexvii
levels
included in the CC calculation for the (e~+~\fexviii) $\rightarrow
$\fexvii system
are:$1s^22s^22p^5 (^2P^o_{1/2,3/2}), 1s^22s^22p^q,n\ell, [S_iL_iJ_i] \
(p=4,5, \ n \leq 4, \ell \leq 3)$. The Rydberg series of
AI resonances correspond to $(S_iL_iJ_i) \ n \ell, \ n \leq 10, \ell
\leq 9$, with effective quantum number defined as a continuous variable
$\nu_i = z/\sqrt(E_i-E) \ (E>0)$, throughout the energy range up to the
highest
$218^{th}$ \fexviii core level; the $n=2,3,4$ core levels range from
E=0-90.7 Ry (\cite{n11,np16}).
 The \fexvii BPRM calculations were carried out resolving the bound-free
cross sections at $\sim$40,000 energies for 454
bound levels with AI resonance structures (in total 587 bound levels are
considered, but the higher lying levels are included to ensure
convergence and completeness as discussed in paper P4,
and do not significantly contribute to
opacities calculations).
Given 217 excited core levels of \fexviii, convolution
is carried out at each energy or approximately $10^9$ times for each
(T,$N_e$) pair.

\begin{figure}
\begin{center}
{ \includegraphics[width=3.0in,height=3.5in]{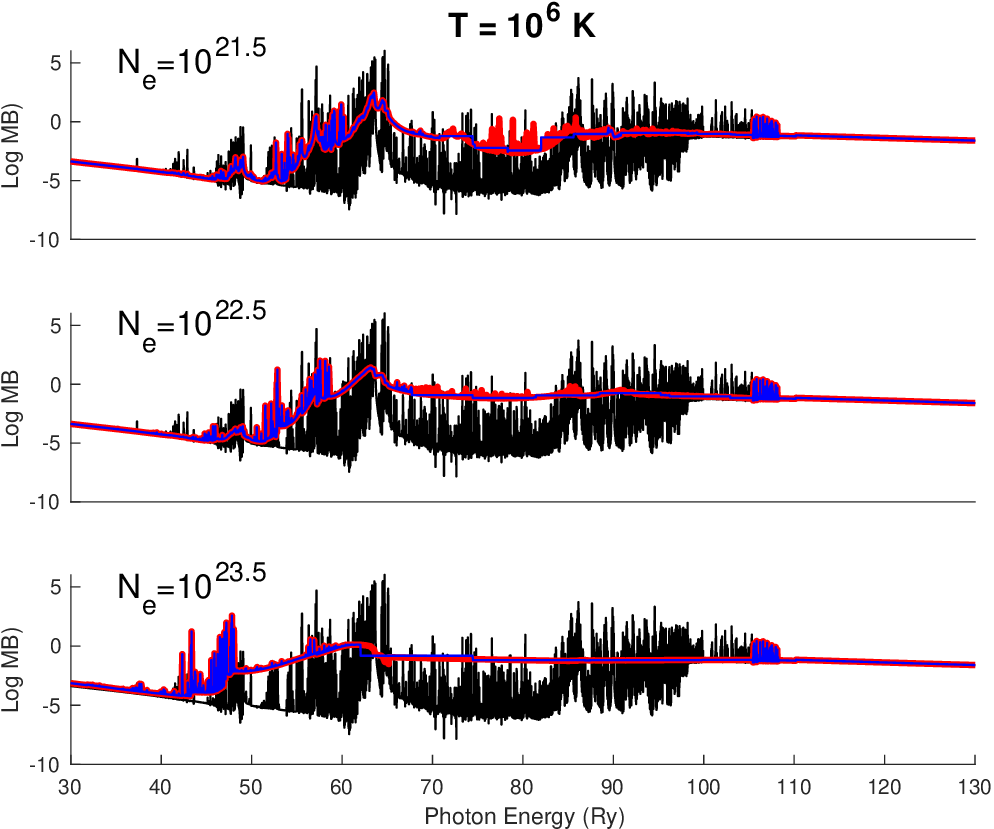}}
{ \includegraphics[width=3.0in,height=3.5in]{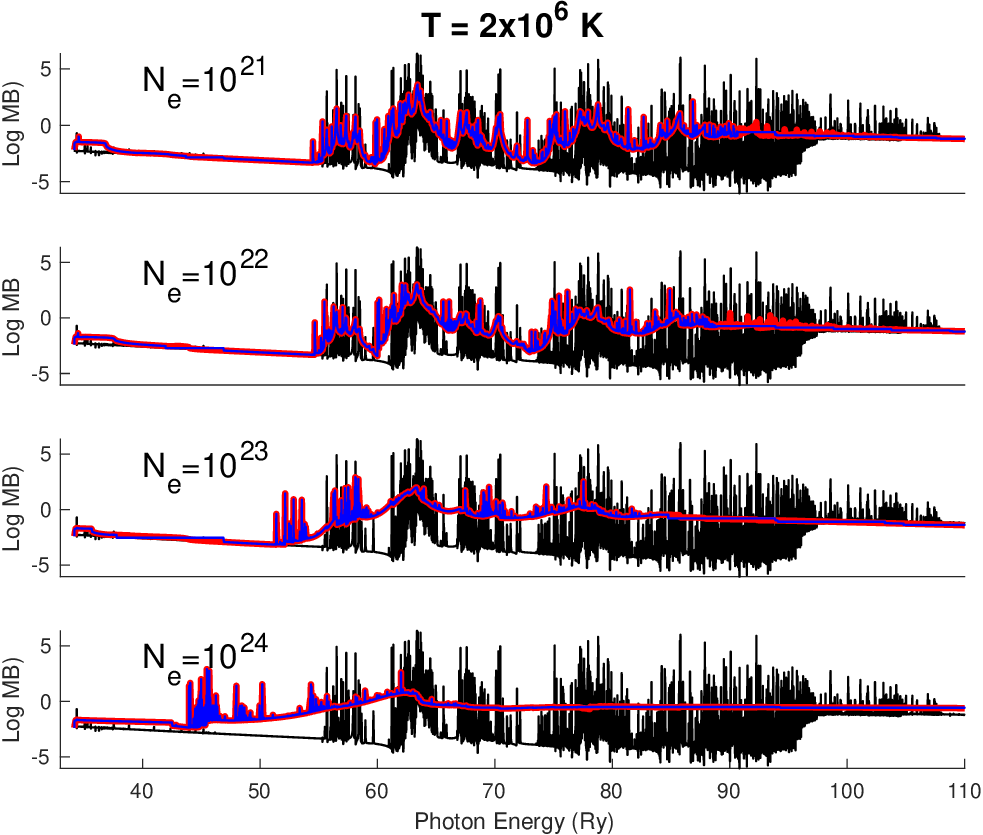}}
\end{center}
\caption{Plasma broadened photoionization cross sections
for $\hbar \omega + \fexvii \rightarrow e~+~\fexviii $ of the bound
level $2s^22p^5[^2P^o_{3/2}]4d(^1F^o_3)$ (left, ionization energy
17.626 Ry), 
level $2s^22p^5[^2P^o_{3/2}]3p (^3D_2)$ (right, ionization energy 37.707
Ry) along two isotherms
$T=1 \times 10^6$K (left) and $T=2 \times 10^6$K (right), 
and electron densities as shown in each panel: black
---
unbroadened, red --- broadened, blue --- broadened with Stark
ionization cut-off $\nu_s^*$ (Table 1). Rydberg series of
AI resonance complexes with $\nu_i \leq 10$ belonging to 217 excited
\fexviii levels $E_i$ broaden and shift with increasing density, also
 resulting in continuum raising and threshold lowering.
\label{fig:fe17}}
\end{figure}

Fig.~\ref{fig:fe17} (left) displays detailed results for plasma broadened and unbroadened
photoionization cross section
of one particular excited level $2s^22p^5[^2P^o_{3/2}]3p (^3D_2)$
(left, ionization energy 37.707 Ry)
of \fexvii along isotherm $T= 10^6$K at three
representative densities (note the $\sim$10 orders of magnitude
variation in resonance heights along the Y-axis). The main feature
evident in the figure are as follows.
(i) AI resonances begin to show significant broadening and smearing of
a multitude of overlapping Rydberg series at
$N_e = 10^{21}$cc. The narrower high-\en \el resonances dissolve into
the
continua but stronger low-\en \el resonance retain their asymmetric
shapes with attenuated heights and widths. (ii) As the density
increases by one to two order of magnitude, to $N_e=10^{22-23}$cc,
resonance
structures not only broaden but their strengths shift and redistributed
over a wide range determined by total width
$\gamma(\omega,\nu_i,T,N_e)$ at each energy $\hbar \omega$ (Eq.~6).
(iii) Stark ionization cut-off (Table 1) results in step-wise structures
that represent the average due to complete dissolution into continua.
(iv) The total AI resonance strengths are conserved, and
integrated values generally do not deviate by more than 1-2\%.
For example, the three cases in Fig.~\ref{fig:fe17} (left):
unbroadened structure (black), and broadened without (red) and with
Stark
cut-off (blue), the integrated numerical values are 59.11, 59.96, 59.94
respectively. This is also an important accuracy
check on numerical integration and the computational algorithm,
as well as the choice of the parameter $\delta$ that determines the
energy range of the Lorentizan profile at each T and $N_e$; in the
present
calculations it varies from $\delta$ = 0.01-0.05 for
\dne=$10^{21-24}$cc.

Fig.~\ref{fig:fe17} (right) shows
similar results to Fig.~\ref{fig:fe17} (left) for another excited \fexvii level
$2s^22p^5[^2P^o_{3/2}]4d(^1F^o_3)$ (ionization energy
17.626 Ry), along a higher
temperature $2 \times 10^6$K isotherm at different intermediate densities. Both
Figs.~2 and 3
show a redward shift of low-\en resonances and dissolution of high-\en
resonances. In addition, the background continuum is raised owing to
redistribution of resonance strengths, which merge into one across high
lying and overlapping thresholds. 

\subsection{\fexviii: Scaling and delineation of resonances}

 Next, we employ plasma broadened cross sections for \fexviii to
highlight the scale, shape, scope, width and magnitude of AI
resonances. 

\begin{figure}
\begin{center}
{ \includegraphics[width=3.0in,height=3.5in]{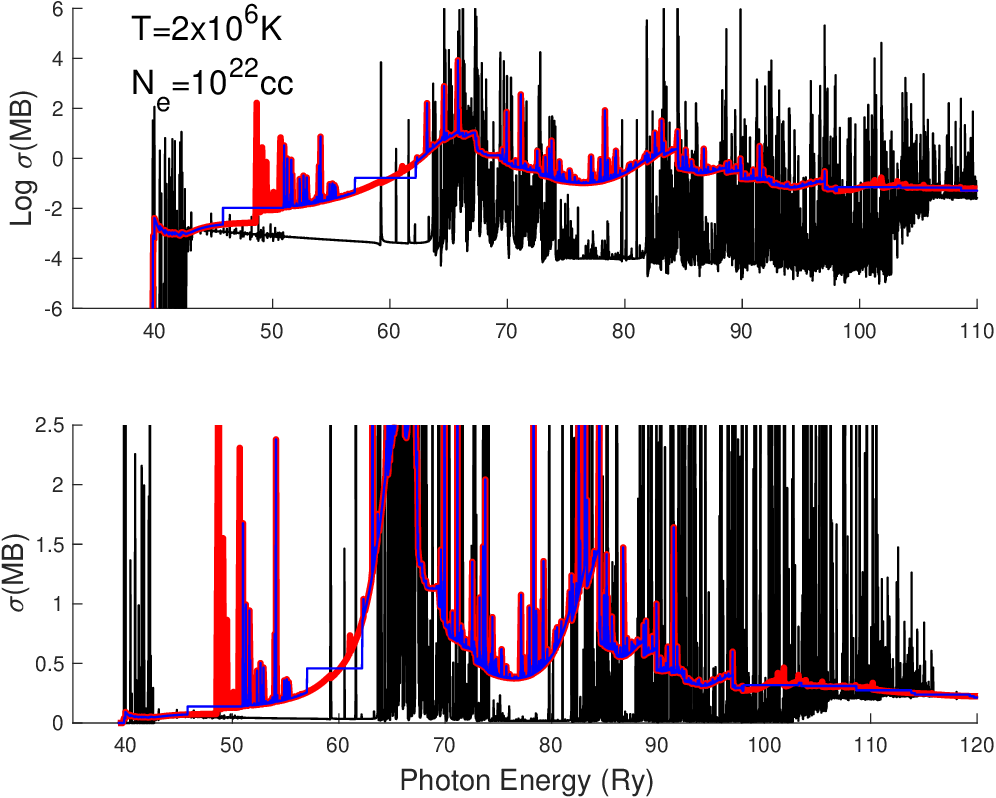}}
{ \includegraphics[width=3.0in,height=3.5in]{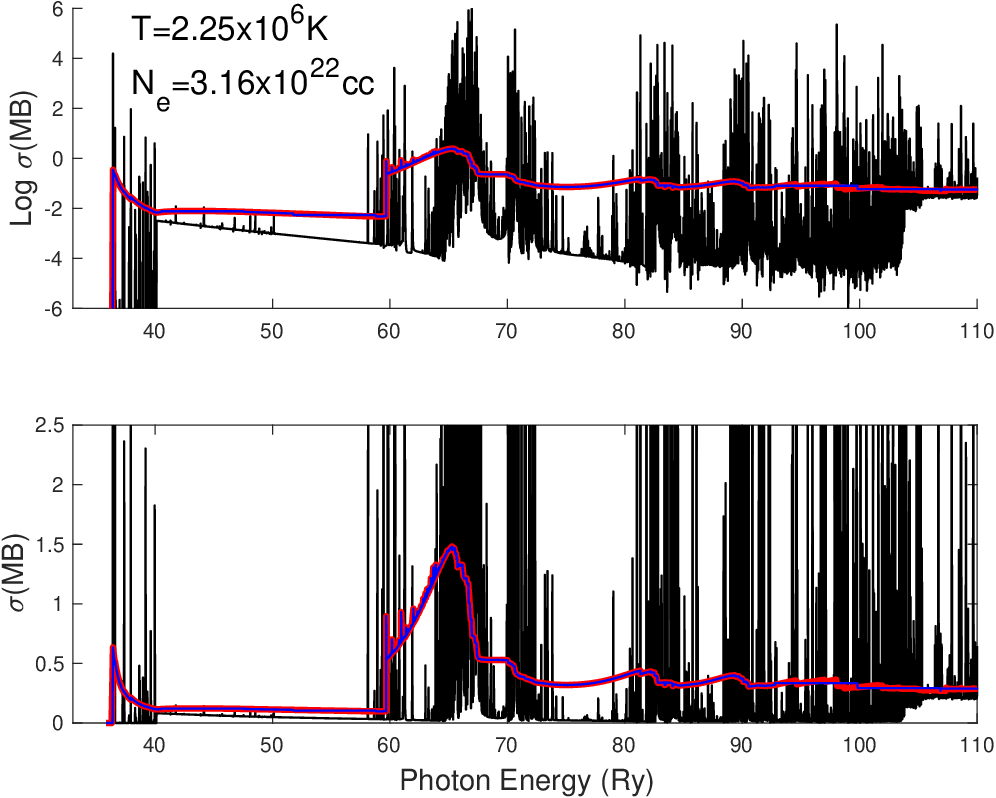}}
\end{center}
\vskip 0.25in
\caption{Plasma broadened photoionization cross sections
on Log and linear scales, $\sigma_{PI}$(MB) (top panels)
and $Log \sigma_{PI}$(MB) (bottom panels)
for $\hbar \omega + \fexviii \rightarrow e~+~\fexix $ of the bound
level $2s^2p^5 \ ^2P^o_{1/2}$ (left, ionization energy 98.903 Ry), and
level $2s^22p^4 (^1D^e_2) 3p \ ^2F^o_{5/2}$ 
(right, ionization energy 39.1204 Ry): black
--- unbroadened, red --- broadened, blue --- broadened with Stark
ionization cut-off $\nu_s^*$ (Table 1). Rydberg series of
AI resonance complexes with $\nu_i \leq 10$ belonging to 276 excited
\fexix levels.
\label{fig:fe18}}
\end{figure}

 The scale of unbroadened AI features is evident upon a comparison on
log and linear scales as in Fig.~\ref{fig:fe18} (black curves),
considered for two excited \fexviii levels. 
The top and bottom
panels on left and right
exhibit $Log \sigma_{PI}$(MB) and $\sigma_{PI}$(MB) respectively.
Whereas the log-scale in top panels appropriately displays the full extent 
of AI resonances, it appears with equal weight for both positive values
that rise up to $10^6$ MB, and for negative values down to
$10^{-6}$ MB that are not significant contributors, as 
shown in the bottom panels on a much smaller linear scale going from
zero only up to 2.5 MB. 

 Attenuation of AI features due to plasma effects are shown in red and
blue curves at two different T-D pairs; cross sections on the left are at a
lower temperature and more than three times lower electron density than
the ones on the right. Consequently, the AI features on the right in
Fig.~\ref{fig:fe18} are much more broadened that the ones on the left.
Two other noticeable features are the closing of "opacity windows" in the
unbroadened cross sections, and shift of AI resonances leading to
temperature-density dependent
redistribution of differential oscillator strengths and opacity with
energy. 

\subsection{Conservation of differential oscillator strength}

 It is important to ensure the numerical accuracy of AI plasma
broadening in temperature-density-energy space. 
Theoretically and computationally, that implies an investigation of integrated
differential oscillator strengths proportional to $\sigma_{PI}$
for all levels of a given ion for the three forms computed: (i)
unbroadened (black curves), (ii) 
with all plasma broadening effects included as in Eq.~(6)
(red curves), and (iii) as in (ii) but with Stark ionization cut-off 
that leads to sharp step-wise structures below each ionization threshold
(blue curves). We had quoted these values for one level of \fexvii
above in Fig.~\ref{fig:fe17}. 

\begin{figure}
\begin{center}
{ \includegraphics[width=3.0in,height=3.5in]{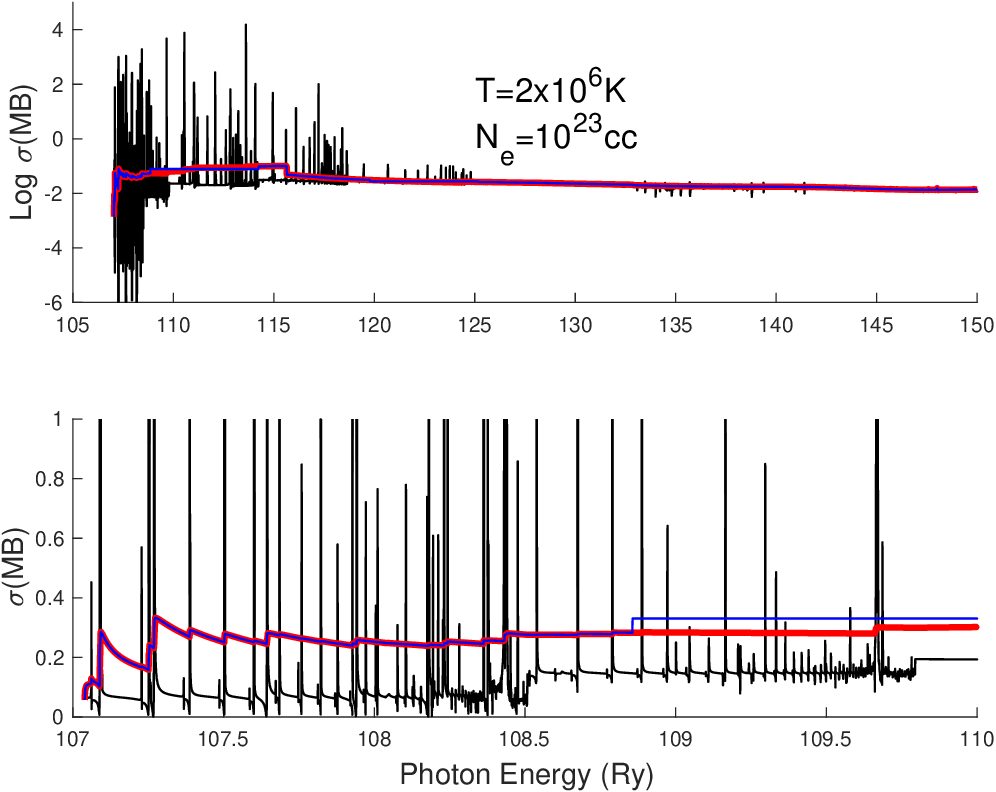}}
{ \includegraphics[width=3.0in,height=3.5in]{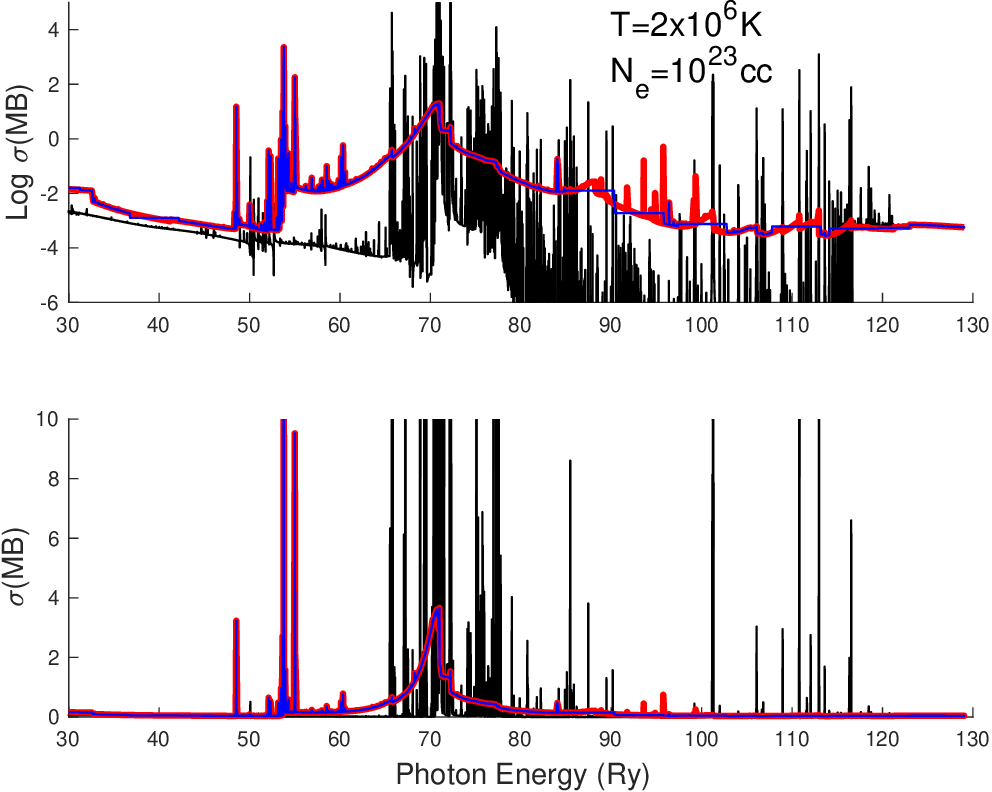}}
\end{center}
\caption{Plasma broadened photoionization cross sections
on Log and linear scales, $\sigma_{PI}$(MB) (top panels)
and $Log \sigma_{PI}$(MB) (bottom panels)
for $\hbar \omega + \fexix \rightarrow e~+~\fexx $ of the ground
state $2s^2p^4 \ ^3P^e$ (left), and $2s2p^4(^2S)3s \ ^1S^e$
(right): black
--- unbroadened, red --- broadened, blue --- broadened with Stark
ionization cut-off $\nu_s^*$ (Table~2). AI resonances in the
unbroadened $\sigma_{PI}$ on the right range over 20 orders of magnitude.
\label{fig:fe19}}
\end{figure}

 In Fig.~\ref{fig:fe19} we present
$\sigma_{PI}$ for the ground state of \fexix $2s^2p^4 \ ^3P^3$
(ionization energy 104.956 Ry), as well as an excited
state $2s2p^4(^2S)3s \ ^1S^e$ (ionization energy 24.186 Ry). 
For these two cross sections of \fexix we find integrated values over the
entire energy range shown to be 21.74, 22.98 and 22.90 for the
unbroadened, broadened, and broadened with Stark ionization cut-off,
respectively for the ground state, and 12.48, 13.57 and 13.56
respectively for the excited state (units are in MB-Ry though only the
relative values are indicators of accuracy). The numerical agreement between the
three sets of values is well within $\sim$10\% indicating conservation
of oscillator strength, despite some uncertainty in integration over
extensive narrow and broad resonance
structures that vary by nearly 20 orders of magnitude in height for
$\sigma_{PI}$($2s2p^4(^2S)3s \ ^1S^e$), and widely disparate width
distribution among Rydberg vs. Seaton PEC resonances described in RMOP2. 

 Generally, the agreement between the three sets of calculations for
each level of each ion at each temperature-density is also an accuracy 
check of the plasma
broadening treatment presented. 
Since there are hundreds of levels or
each ion considered, there is more than 10\% difference in integrated
cross sections for highly excited levels at very high densities where the
total AI width (Eq.~6) is very large. However, the highly excited
levels are cut-off by the MHD-EOS and not considered in opacity calculations. 

\subsection{Plasma opacity parameters}

 Table~2 gives plasma parameters corresponding to Figs.~\ref{fig:fe17}.
Their physical significance is demonstrated by a representative sample
tabulated temperature T(K) and \dne. The maximum
width $\gamma_{10}$ corresponding to $\nu_i=10$ in Eqs. (3,6) is set
by the CC-BPRM calculations which delineate unbroadened AI resonance
profiles up to $\nu \leq 10$, and employ an averaging procedure up to
each
threshold $ 10 < \nu < \infty$ using quantum defect (QED)
theory (viz. \cite{nb3,op,aas} and references therein).
$\gamma_c(10)$ and $\gamma_s(10)$ are the maximum collisional and Stark
width components. The Doppler width $\gamma_d$ is much smaller,
$1.18\times 10^{-3}$ and $1.67\times 10^{-3}$ Ry at $10^6$K and $2\times
10^6$K respectively, validating its inclusion in Eq.~(6) in HED plasma
sources but possibly not when $\gamma_d$ is
comparable to $\gamma_c$ or $\gamma_s$. The $\nu^*_s$ and
$\nu_D$ are effective quantum numbers corresponding to Stark
ionization cut-off and the Debye radius respectively. We obtain
$\nu_D = \left[ \frac{2}{5}\pi z^2 \lambda_D^2 \right ]^{1/4}$,
where the Debye length $\lambda_D = (kT/8\pi N_e)^{1/2}$.
It is seen in Table~2 that
$\nu_D > \nu^*_s$ at the T, \dne considered, justifying neglect of
plasma screening effects herein,
but which may need to be accounted for at even higher
densities.

 The aggregate effect of AI broadening for large-scale applications
is demonstrated in Table~2 by the
ratio R of the Rosseland Mean
Opacity (discussed in paper RMOP1)
using broadened/unbroadened cross sections for 454 \fexvii
levels with AI resonances (other higher bound levels have negligible
resonances) \cite{nb4,np16}. For any atom or ion R
is highly dependent on T and \dne; for \fexvii R yields up to 58\%
enhancement due to plasma broadening with increasing
\dne along the $2\times 10^6$K isotherm,
but decreasing to 6\% along the $10^6$K isotherm. Approximately 70,000
free-free transitions among +ve energy levels are included in the
calculation of
R, but their contribution has no significant broadening effect since
they entail very high-lying levels with negligible level populations.
However, different plasma environments with intense radiation fields, or
a different equation-of-state than \cite{mhd} employed here, may lead to
more discernible effect due to free-free transitions. AI broadening
in a plasma environment
affects each level cross section differently, and hence
its contribution to opacities or rate equations for atomic processes
in general. A critical (T,\dne) range can therefore be numerically
ascertained where redistribution and shifts of atomic resonance
strengths would be significant and cross sections should be modified.

\begin{table}
\caption{Plasma parameters along isotherms in Fig.~2 and
3; $\nu_D$ corresponds to Debye radius; R is the ratio of \fexvii
Rosseland Mean Opacity with and without broadening
\cite{nb4}; $\gamma_{10}$ is the maximum AI resonance width at
$\nu=10$.}
\begin{center}
\begin{tabular} {c|c|c|c|c|c|c|c}
\hline
T(K) & $ N_e (cc)$ & $\gamma_{10}(Ry)$ & $\gamma_c(10)$ & $\gamma_s(10)$
&
$\nu_s^*$ & $\nu_D$ & R \\
& & $\nu=10$ & & & & &\\
\hline
 $2 \times 10^6$ & $10^{21}$ & 3.42(-1)& 8.55(-2) & 2.57(-1) & 10.4 &
28.1& 1.35 \\
 $2 \times 10^6$ & $10^{22}$ & 2.05(0) & 8.55(-1) & 1.19(0) & 7.7 &
15.8 & 1.43 \\
 $2 \times 10^6$ & $10^{23}$ & 1.41(1) & 8.55(0) & 5.53(0) & 5.6 & 8.9 &
1.55 \\
 $2 \times 10^6$ & $10^{24}$ & 1.11(2) & 8.55(1) & 2.57(1) & 4.1 & 5.0 &
1.58 \\
 $10^6$ & $3.1\times10^{21.5}$ & 8.17(-1) & 2.71(-1)& 5.46(-1)& 9.0 &
17.8 & 1.47 \\
 $10^6$ & $3.1\times10^{22.5}$ & 5.25(0) & 2.71(0) & 2.53(0)& 6.6 & 10.0
& 1.13 \\
 $10^6$ & $3.1\times10^{23.5}$ & 3.89(0) & 2.71(1)& 1.18(0) & 4.8 & 5.6
& 1.06 \\
\hline
\end{tabular}
\end{center}
\end{table}

\section{Conclusion}

 The main conclusions are: (I) The method described herein is generally
applicable to AI resonances in atomic processes in HED plasmas. (II)
The cross sections become energy-temperature-density dependent in a
The cross sections become energy-temperature-density dependent in a
critical range leading to broadening, shifting, and dissolving into
continua. (III) Among the approximations
necessary to generalize the formalism is the assumption that thermal
Doppler widths are small compared to collisional and Stark widths as
herein, but given the intrinsic asymmetries of AI resonances it may not
lead to significant inaccuracies (although that needs to be verified in
future works). (IV) The treatment of Stark broadening and ionization
cut-off is {\it ad hoc}, albeit based on the equation-of-state
formulation \cite{mhd} and consistent with previous works \cite{op}.
(V) Since it is negligibly small,
the free-free contribution is included post-facto
in the computation of the ratio R in Table~2 and not in the
cross sections and results shown in Figs.~2 and 3, but may be
important in special HED environments with intense radiation and should
then be incorporated in the main calculations of total AI width (Eq.~6).
(VI) The predicted redward shift of AI resonances as the plasma
density increases should be experimentally verifiable.
(VII) Redistribution of AI resonance strengths should particularly
manifest itself in rate coefficients for
\eion excitation and recombination in plasma models and simulations, and
for
photoabsorption in opacity calculations, using temperature-dependent
Maxwellian, Planck, or other particle distribution functions.
(VIII) The treatment of
individual contributions to AI broadening may be improved,
and the theoretical formulation outlined
here is predicated on the assumption that external plasma effects are
perturbations subsumed by and overlying the intrinsic autoionization
effect.
(IX) The computational formalism is designed to be amenable for
practical
applications and
the computational algorithm and a general-purpose program AUTOBRO
are optimized for large-scale computations
of AI broadened cross sections for atomic processes 
in HED plasma and astrophysical models. 

\vskip 0.25in
{\bf Acknowledgments}
\vskip 0.25in

 I would like to thank Sultana Nahar for atomic data for Fe ions and
discussions.


\begin{thebibliography}{10}
\bibitem{nb1} Several formulations were investigated and
employed to implement a Voigt
profile used in line broadening work (viz. \cite{h65,h78,h04} and later works).
They were found to be numerically unstable owing to required limits 
on the relative widths of AI broadening mechanisms considered herein, 
that rendered the
Voigt kernel mathematically ill-conditioned and computationally
impractical. For example, the parameter related to the ratio 
collisional-to-Doppler widths $\gamma_c/\gamma/_d$, that is
usually evaluated and specified for line broadening and assigned
numerical limiting values in integration, is not practical since
$\gamma_c >> \gamma_d$ at temperatures and densities of interest in HED
plasmas.
\bibitem{nb3} The QD regions $10 < \nu_i < \infty$ are
generally small and overlapping for closely spaced excited core
levels, as herein.
\end{thebibliography}
\end{document}